\documentclass[numeric]{aa}
\usepackage[dvipsnames]{xcolor}

\usepackage{graphicx}
\usepackage{txfonts}
\usepackage{hyperref}

\begin{document} 

\newcommand{\afffias}{Frankfurt Institute for Advanced Studies (FIAS), Ruth-Moufang-Strasse~1, 60438 Frankfurt am Main, Germany}
\newcommand{\affbgu}{Physics Department, Ben-Gurion University of the Negev, Beer-Sheva 84105, Israel}
\newcommand{\affbul}{Institute for Nuclear Research and Nuclear Energy, Bulgarian Academy of Sciences, Sofia, Bulgaria}

   \title{ {Testing Late-Time Cosmic Acceleration with uncorrelated Baryon Acoustic Oscillations dataset}}

   \author{David Benisty\inst{1,2}
          \and
          Denitsa Staicova\inst{3}}

   \institute{
$^{1}\,$ Frankfurt Institute for Advanced Studies (FIAS), Ruth-Moufang-Strasse~1, 60438 Frankfurt am Main, Germany \\  
$^{2}\,$ Physics Department, Ben-Gurion University of the Negev, Beer-Sheva 84105, Israel
\\$^{3}\,$ Institute for Nuclear Research and Nuclear Energy, Bulgarian Academy of Sciences, Sofia, Bulgaria}
   \date{}

 
  \abstract{ {Baryon Acoustic Oscillations (BAO) involve measuring the spatial distribution of galaxies to determine the growth rate of cosmic structure. We derive constraints on cosmological parameters from $17$ uncorrelated BAO measurements that were collected from $333$ published data points in the effective redshift range $0.106 \leq z \leq 2.36$. We test the correlation of the subset using random covariance matrix. The $\Lambda$CDM model fit yields the cosmological parameters: $\Omega_m = 0.261 \pm 0.028$ and $\Omega_\Lambda = 0.733 \pm 0.021$. Combining the BAO data with the Cosmic Chronometers data, the Pantheon Type Ia supernova and the Hubble Diagram of Gamma Ray Bursts and Quasars, the Hubble constant yields $ 69.85 \pm 1.27 km/sec/Mpc$ and the sound horizon distance gives: $ 146.1 \pm 2.15 Mpc$. Beyond the $\Lambda$CDM model we test $\Omega_K$CDM and wCDM. The spatial curvature is $\Omega_k = -0.076 \pm 0.012$ and the dark energy equation of states: $w = -0.989 \pm 0.049$.  {We perform AIC test to compare the 3 models and see that $\Lambda$CDM scores best.}}
}
  


%

\maketitle
\section{Introduction}
The Standard model of cosmology, the $\Lambda$CDM model, requires a dark energy (DE) component responsible for the observed late-time acceleration of the expansion rate. The tension between the values of the Hubble constant $H_0$ obtained from the late universe measurements (\cite{Riess:2019cxk}) and those from
the Cosmic Microwave Background (CMB) by Planck Collaboration (\cite{Aghanim:2018eyx}) is larger than $4\sigma$. This tension is one of the biggest challenges in modern cosmology (\cite{DiValentino:2020vhf,DiValentino:2020zio,DiValentino:2020vvd,Efstathiou:2020wxn,Borhanian:2020vyr,Hryczuk:2020jhi,Klypin:2020tud,Ivanov:2020mfr,Chudaykin:2020acu,Lyu:2020lps,Alestas:2020mvb,Motloch:2019gux,Frusciante:2019puu}).

The measurement of the expansion history of the Universe
at low redshifts provides observational tests for the dark energy while counting on  different types of astrophysical objects and observational techniques. The use of type Ia supernovae (SNe) as Standard candles originally established the accelerated expansion and solidified the introduction of the DE (\cite{Perlmutter:1998np,Riess:1998cb}). Baryon Acoustic Oscillations (BAO) provide a Standard ruler which has been evolving with the Universe since the recombination epoch (\cite{Cuceu:2019for,Wu:2020nxz}). The BAO scale at different times is a powerful tool to constraint the cosmological parameters. BAO are present in the distribution of matter and any tracer of the large-scale structures can be used to measure the BAO peak. Current surveys include measurements at different redshifts using the clustering of galaxies (emission-line galaxies -- ELG and luminous red galaxies -- LRG) and quasars, from the correlation function of the Ly$\alpha$ absorption lines in the spectra of distant quasars, in cross correlation with quasar positions and galaxies, in Fourier and configuration space. Their biggest attraction is the ability to study the cosmic acceleration independently from the Supernova and the CMB surveys and thus to give a possible answer to the different observed tensions. This has been done already with different datasets (\cite{Handley:2019tkm,DiValentino:2020hov,Luo:2020ufj}).

 { {In this article, we gather the largest collection of BAO data points (333 points) and from it we select 17 uncorrelated BAO points.}  {This is because using} the complete collection of BAO data (333 data points) would lead to a very large error due to data correlations. Therefore we select a smaller dataset which may reduce the errors. We test the stability of this set against the inclusion of random correlations and then we perform standard analysis on the cosmological parameters.}

 {The paper is structured as follows: Section \ref{sec:back} describes the theoretical background. Section \ref{sec:meth} describes the methodology with the cross correlation validation. Section \ref{sec:LCDM} discusses the $\Lambda$CDM fit and section \ref{sec:ext} discusses on the extensions. Finally, section \ref{sec:Dis} summarizes our results.}

\begin{table*} 
\scalebox{1.1}{
\begin{tabular}{ccccccc}       
\hline\hline                                                                                                    

$z$ & Parameter   & Value & Error & year  & Survey &  Ref. \\ \hline\hline 
$0.106$& $r_d/D_V$ & $0.336$& $0.015$&  $2011$ & 6dFGS & \cite{Beutler:2011hx}\\
$0.15$& $D_V (r_{d, fidd}/r_d)$ & $664$& $25.0$&  $2014$ & SDSS DR7 & \cite{Ross:2014qpa}\\
$0.275$& $r_d/D_V$& $0.1390$& $0.0037$&   $2009$& SDSS-DR7+2dFGRS& \cite{Percival:2009xn}\\
$0.32$& $D_V (r_{d, fidd}/r_d)$ & $1264$& $25$&  $2016$& SDSS-DR11 LOWZ& \cite{Tojeiro:2014eea}\\
$0.44$& $r_d/D_V$ & $0.0870$& $0.0042$&  $2012$& WiggleZ& \cite{Blake:2012pj}\\
$0.54$& $D_A/r_d$ & $9.212$& $0.41$&  $2012$  &SDSS-III DR8& \cite{Seo:2012xy}\\
$0.57$&  $D_V/r_d$& $13.67$& $0.22$& $2012$ &SDSSIII/DR9& \cite{Anderson:2012sa}\\
$0.6$& $r_d/D_V$ & $0.0672$ & $0.0031$& $2012$ & WiggleZ& \cite{Blake:2012pj}\\
$0.697$& $D_A (r_{d, fidd}/r_d)$ & $1499$ & $77$ & $2020$ & DECals DR8 & \cite{Sridhar:2020czy}\\
$0.72$& $D_V (r_{d, fidd}/r_d)$ & $2353$& $63$& $2017$ &SDSS-IV DR14& \cite{Bautista:2017wwp}\\
$0.73$& $r_d/D_V$ & $0.0593$& $0.0020$& $2012$ & WiggleZ & \cite{Blake:2012pj}\\
$0.81$& $D_A/r_d$ & $10.75$& 0.43&   $2017$ &DES Year1& \cite{Abbott:2017wcz}\\
$0.874$& $D_A (r_{d, fidd}/r_d)$ & $1680$ & $109$& $2020$ & DECals DR8 & \cite{Sridhar:2020czy}\\
$1.48$& $D_H \cdot r_d $& $13.23$ & $0.47$ &  $2020$ & eBoss DR16 BAO+RSD & \cite{Hou:2020rse}\\
$1.52$& $D_V (r_{d, fidd}/r_d)$ & $3843$& 147.0&  $2017$ & SDSS-IV/DR14 & \cite{Ata:2017dya}\\
$2.3$& $H \cdot r_d $& $34188$ & $    1188$ &  $2012$ &Boss Lya quasars DR9& \cite{Busca:2012bu}\\
$2.34$& $D_H \cdot r_d $& $8.86$ & $0.29$ &  $2019$ &BOSS DR14 Lya in LyBeta & \cite{Agathe:2019vsu}\\
\hline\hline                                                  
\end{tabular} 
}
\caption{ {\it{The uncorrelated dataset that has been used is this paper. For each redshift, the table presents the parameter, the mean value and the corresponding error bar. The Ref. and the collaboration is also reported.}}}
\label{tab:data}                                              
\end{table*}                                               
\section{Theoretical Background}
\label{sec:back}
We assume a Friedmann-{Lema\^itre}-Robertson-Walker metric with the scale parameter $a = 1/(1+z)$, where $z$ is the redshift. The Friedmann equation for $\Lambda$CDM background reads:  
\begin{equation}
    E(z)^2 = \Omega_{r} (1+z)^4 + \Omega_{m} (1+z)^3 + \Omega_{k} (1+z)^2 + \Omega_{\Lambda},
    \label{eq:hzlcdm}
\end{equation}
where $\Omega_{r}$, $\Omega_{m}$, $\Omega_{\Lambda}$ and $\Omega_{k}$ are the fractional densities of radiation, matter, dark energy and the spatial curvature at redshift $z=0$. The function $E(z)$ is the ratio $H(z)/H_0$, where $H(z) := \dot{a}/a$ is the Hubble parameter at redshift $z$ and $H_0$ is the Hubble parameter today. The radiation density  can be computed as: $\Omega_r = 1 - \Omega_m - \Omega_\Lambda - \Omega_{k}$. In the late universe, at the redshifts probed by BAO, the radiation fraction is very small, while for a flat Universe, $\Omega_k=0$. For wCDM, the Friedmann equation is generalized to $\Omega_{\Lambda}\to \Omega_{DE}^{0}  (1+z)^{-3(1+w)}$.

The observed tracer redshifts and angles on the sky need to be converted to distances by adopting a fiducial cosmological model, and the analysis measures the ratio of the observed BAO scale to that predicted in the fiducial model.

Studies of the BAO feature in the transverse direction provide a measurement of $D_H(z)/r_d = c/H(z)r_d$, with the comoving angular diameter distance (\cite{Hogg:2020ktc,Martinelli:2020hud}): 
\begin{equation}
D_M=\frac{c}{H_0} S_k\left(\int_0^z\frac{dz'}{E(z')}\right), 
\end{equation}
where 
 {\begin{equation}
S_k(x) = 
\begin{cases}
\frac{1}{\sqrt{\Omega_k}}\sinh\left(\sqrt{\Omega_k}x\right) \quad &\text{if}\quad \Omega_k>0
\\
x \quad  &\text{if} \quad \Omega_k=0  
\\
\frac{1}{\sqrt{-\Omega_k}}\sin\left(\sqrt{-\Omega_k} x\right)\quad &\text{if} \quad \Omega_k<0.
\end{cases}
\end{equation}}
In our database we also use the angular diameter distance $D_A=D_M/(1+z)$ and the $D_V(z)/r_d$, which is a combination of the BAO peak coordinates above:
\begin{equation}
    D_V(z) \equiv [ z D_H(z) D_M^2(z) ]^{1/3}.
\end{equation}
$r_d$ is the sound horizon at the drag epoch and it is discussed in the corresponding section. Finally for very precise  “line-of-sight” (or “radial”) observations, BAO can also measure directly the Hubble parameter (\cite{Benitez:2008fs}). 
\begin{figure*}[t!]
 	\centering
\includegraphics[width=0.8\textwidth]{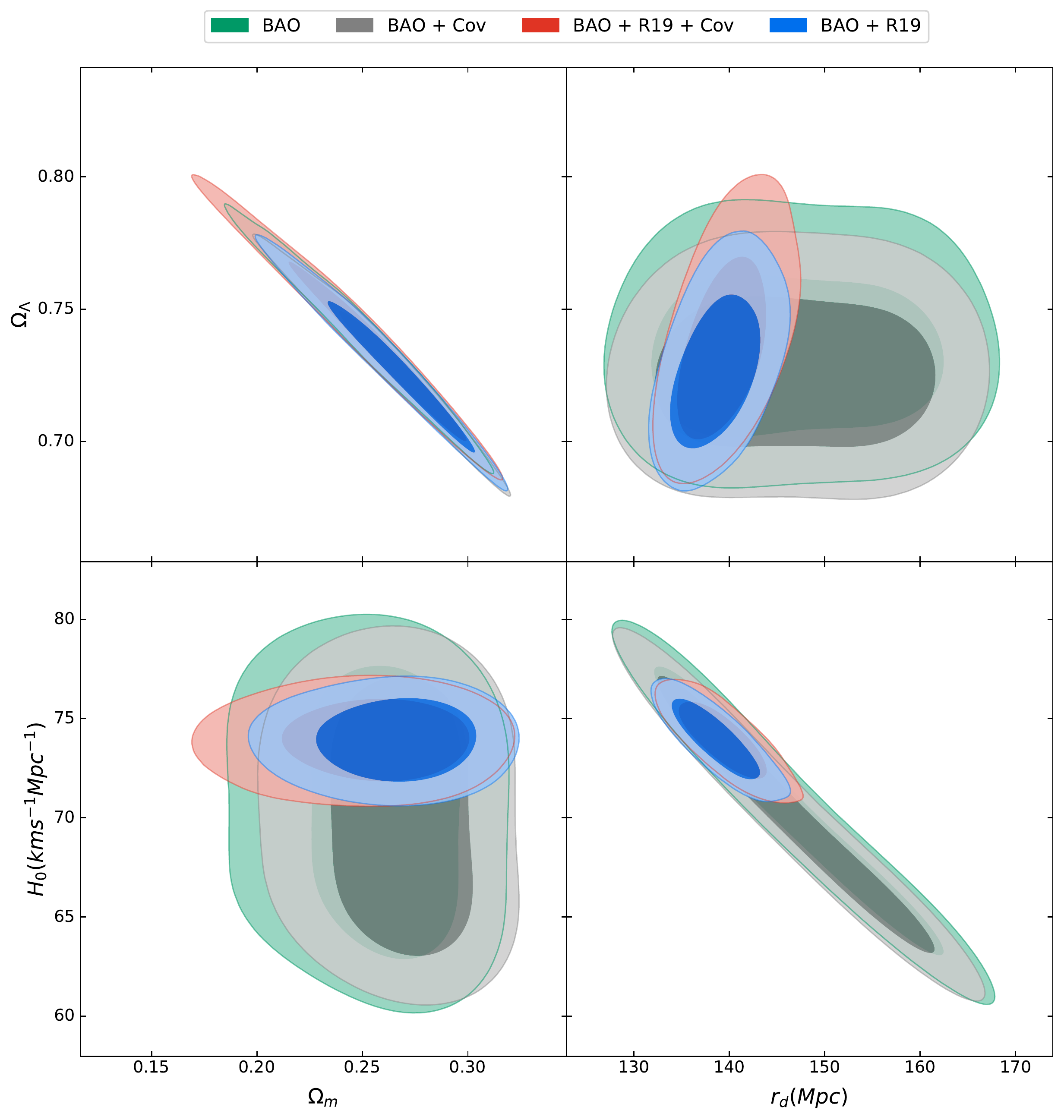}
\caption{\it{ {The posterior distribution for $\Lambda$CDM with and without a test random covariance matrix with $6$ components. The distribution with covariance matrix gives a slightly bigger error bar, but still very similar to the uncorrelated dataset.} }} 
 	\label{fig:check_excerpts}
\end{figure*} 
\begin{table*} 
\centering
\scalebox{1.1}{
\begin{tabular}{ccc}       
\hline\hline                        
n & BAO & BAO + R19 \\
\hline\hline
$n=0$  & $\Omega_m=0.257\pm 0.02$ & $\Omega_m=0.255 \pm 0.03$\\
       & $\Omega_\Lambda= 0.735 \pm 0.021$ & $\Omega_\Lambda= 0.736 \pm 0.021$\\
       & & $r_d=139.32 \pm 2.88\,{Mpc}$\\
\hline
$n=3$ & $\Omega_m=0.268\pm 0.023$ & $\Omega_m=0.267 \pm 0.021$\\
       & $\Omega_\Lambda= 0.725 \pm 0.019$ & $\Omega_\Lambda= 0.725 \pm 0.018$\\
       & & $r_d=138.49 \pm 3.03\,{Mpc}$\\
       \hline
$n=6$ & $\Omega_m=0.275\pm 0.021$ & $\Omega_m=0.274 \pm 0.020$\\
       & $\Omega_\Lambda= 0.719 \pm 0.016$ & $\Omega_\Lambda= 0.720 \pm 0.012$\\
       & & $r_d=138.32 \pm 2.76\,{Mpc}$\\
\hline
\end{tabular} 
}
\caption{\it{ {Variation of the derived parameters depending on the number of correlated points $n$.}}}
\label{tab:test}                                     
\end{table*}  
\section{Methodology}
\label{sec:meth}
 {The dataset we use includes a broad collection of points. The main contribution to our dataset comes from the different data releases (DR) of the Sloan Digital Sky Survey (SDSS): (\cite{Seo:2012xy, Anderson:2012sa,Anderson:2013oza,Moresco:2016mzx, Delubac:2014aqe,Anderson:2013zyy, Gil-Marin:2015nqa,Chuang:2016uuz,Wang:2017wia,Zhao:2016das,Vargas-Magana:2016imr, Cuesta:2015mqa, Beutler:2016ixs, Carvalho:2015ica, deCarvalho:2017xye, Icaza-Lizaola:2019zgk,Bautista:2017wwp, Zhang:2012mp,Chuang:2012qt,Oka:2013cba, Gaztanaga:2008xz, Gil-Marin:2015sqa,Gil-Marin:2016wya, Tojeiro:2014eea, Alam:2016hwk, Ross:2014qpa,Percival:2009xn, Bautista:2017zgn,Chuang:2013wga,Wang:2016wjr, Ata:2017dya,Gil-Marin:2018cgo, Wang:2020tje,Tamone:2020qrl,deMattia:2020fkb,Bautista:2020ahg,Zhao:2020tis, Zhao:2018jxv,Zhao:2018wix,Zhu:2018edv, Gil-Marin:2020bct,Nadathur:2020vld, Raichoor:2020vio, Hou:2020rse,Neveux:2020voa, Busca:2012bu,Bourboux:2017cbm, Agathe:2019vsu,Blomqvist:2019rah,duMasdesBourboux:2020pck, Delubac:2014aqe,Font-Ribera:2013wce}). To these data points, we add the results from the WiggleZ Dark Energy Survey (\cite{Blake:2012pj,Kazin:2014qga}), the Dark Energy Survey (DES) (\cite{Abbott:2017wcz}) and the Dark Energy Camera Legacy Survey (DECaLS) (\cite{Sridhar:2020czy}) (LRG). Furthermore, we use data from the 6dF Galaxy Survey (6dFGS) (\cite{Beutler:2011hx,Carter:2018vce}). We also used some earlier tables of BAO data to improve our datasets  (\cite{Kang:2019vsk,Akarsu:2019pwn,Nesseris:2019fwr,Zhang:2020uan,Benisty:2020kdt}). As a whole we gather 333 BAO points including both the angular distances $D_H$, $D_M$, $D_A$ and $D_V$ and the Hubble parameters multiplied by $r_d$ or $r_d/r_{d, fid}$ where $r_{d, fid}$ is the fiducial sound horizon.}

 {The big collection of data points is highly correlated. Therefore, from this dataset we select an uncorrelated sub-sample on which we perform our analysis.  The final dataset we use is a set of uncorrelated data points from different \textbf{Baryon Acoustic Oscillations (BAO)} measurements from
(\cite{Beutler:2011hx, Ross:2014qpa,Percival:2009xn,Tojeiro:2014eea,Blake:2012pj,Seo:2012xy,Anderson:2012sa,Sridhar:2020czy,Bautista:2017wwp,Abbott:2017wcz,Hou:2020rse,Ata:2017dya,Busca:2012bu,Agathe:2019vsu}).
The collection is summarized in Table \ref{tab:data}.}

A significant problem for the BAO dataset is the possible correlation   {between the different measurements in the diverse data releases}. To evaluate the systematic error, one usually uses mocks based on N-body simulations with known cosmology which produce the appropriate covariance matrices. Since we use a sample of different experiments, we do not use the exact covariance matrix between them, which is not known. Instead we perform covariance analysis based on the one proposed in (\cite{Kazantzidis:2018rnb}). The covariance matrix for uncorrelated points is :
\begin{equation}
C_{ii} = \sigma_i^2.
\end{equation}
To mock the effect of possible correlations among data points, one can introduce a number of randomly selected nondiagonal elements in the covariance matrix while keeping it symmetric. With this approach we introduce positive correlations in up to 6 pairs of randomly selected data points (more than $25\%$ of the data). The positions of the non-diagonal elements are chosen randomly and the magnitude of those randomly selected covariance matrix element $C_{ij}$ is set to
\begin{equation} 
C_{ij} =0.5 \sigma_i \sigma_j
\label{cij}
\end{equation}
where $\sigma_i \sigma_j$ are the published $1\sigma$ errors of the data points $i,j$.

We use a nested sampler as it is implemented within the open-source package $Polychord$ (\cite{Handley:2015fda}) with the $GetDist$ package (\cite{Lewis:2019xzd}) to present the results. The prior we choose is with a uniform distribution, where $\Omega_{m} \in [0.;1.]$, $\Omega_{\Lambda}\in[0.;1 - \Omega_{m}]$, $H_0\in [50;100]$ and $r_d\in [100;200] Mpc$. The measurement of the Hubble constant yielding $H_0 = 74.03 \pm 1.42 (km/s)/Mpc$ at $68\%$ CL by (\cite{Riess:2019cxk}) has been incorporated into our analysis as an additional prior which we denote as {\bf R19}. Note that the prior for $r_d$ is rather wide since the very narrow prior for $r_d$ has an effect of specifying the Hubble parameter (\cite{Knox_2020}) which we tried to avoid in our analysis. With respect to the fiducial cosmology, we use as a prior for the ratio $r_d/r_{d, fid}\in [0.9,1.1]$.

 {The results for the {\bf BAO} and the {\bf BAO} + {\bf R19} can be seen on Fig. \ref{fig:check_excerpts} and in the table below it. As one can see including the correlations changes the results, as expected, but it doesn't lead to a significant difference. The difference between no correlation and 30\% correlated points is about 10\% which is similar to the one in (\cite{Kazantzidis:2018rnb}), but in our case, $r_d$ and the fiducial cosmology are free parameters. Thus we consider that even though this approach may underestimate the covariance matrices, it gives us a justification to consider the points uncorrelated.}

 {In order to constraint the cosmological models, in addition to the BAO dataset we use \textbf{Cosmic Chronometers (CC)} and \textbf{Standard Candles (SC)}. The \textbf{Cosmic Chronometers (CC)} exploit the evolution of differential ages of passive galaxies at different redshifts to directly constrain the Hubble parameter (\cite{Jimenez:2001gg, Stern:2009ep,Moresco:2012jh,Ratsimbazafy:2017vga,Moresco:2015cya,Pellejero-Ibanez:2016ypj}).
The set includes 30 uncorrelated CC measurements of $H(z)$, as discussed in (\cite{Moresco:2012by,Moresco:2012jh,Moresco:2015cya,Moresco:2016mzx}). For \textbf{Standard Candles (SC)} we use uncorrelated measurements of the Pantheon Type Ia supernova dataset (\cite{Scolnic:2017caz}) that were collected in (\cite{Anagnostopoulos:2020ctz}) and the measurements from Quasars (\cite{Roberts:2017nkm}) and Gamma Ray Bursts (\cite{Demianski:2016zxi}).}  {Below, we refer to the dataset including BAO, CC and SC as the \textbf{"full"} dataset.}

\begin{figure*}[t!]
 	\centering
\includegraphics[width=0.9\textwidth]{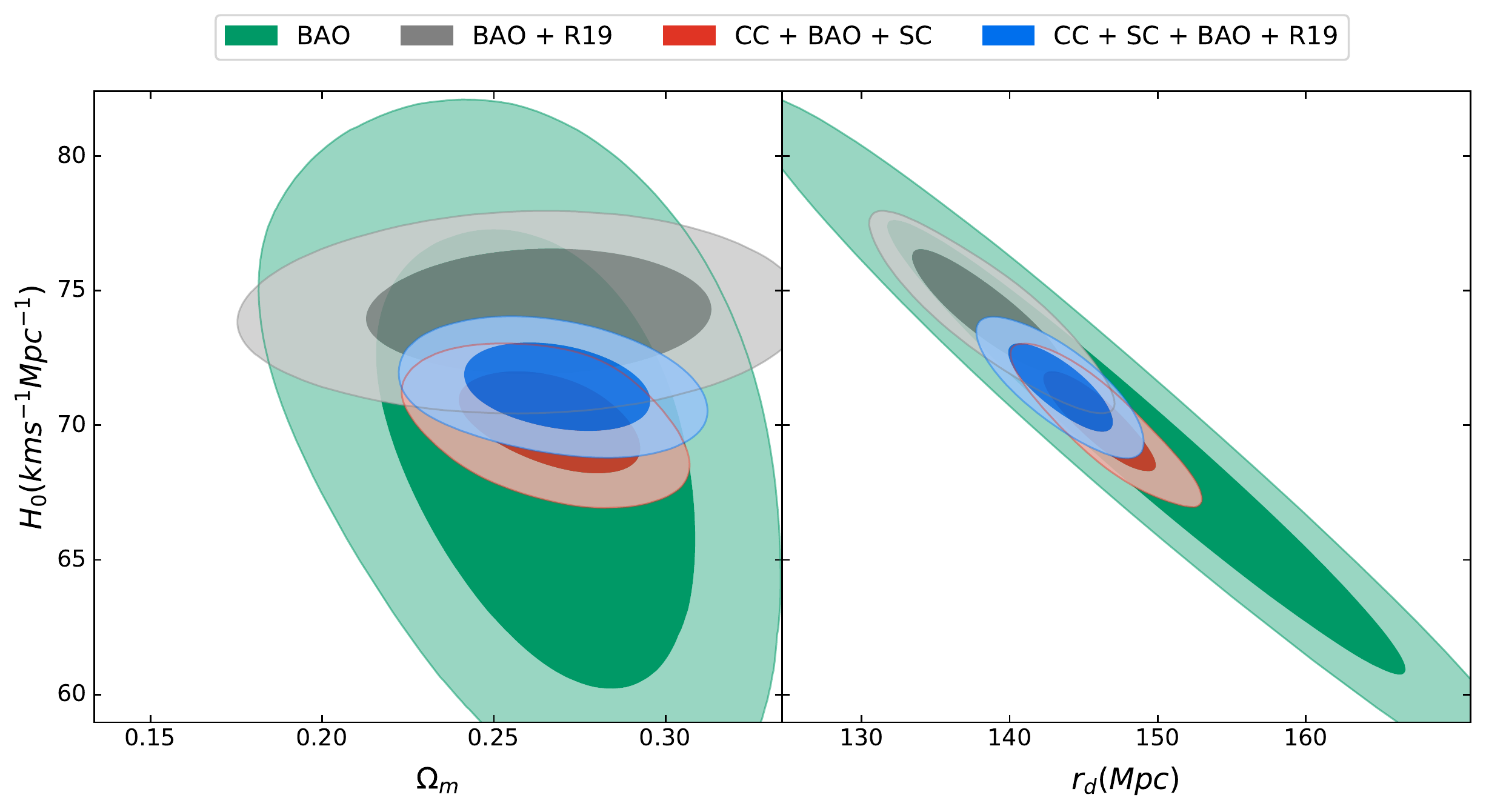}
\caption{\it{The posterior distribution for different measurements with the $\Lambda$CDM model with $1 \sigma$ and $2\sigma$. The BAO refers to the Baryon Acoustic Oscillations dataset from table \ref{tab:data}. The CC dataset refers to Cosmic Chronometers and SC refers to the Hubble Diagram from Type Ia supernova, Quasars and Gamma Ray Bursts. R19 denotes the Riess 2019 measurement of the Hubble constant as a Gaussian prior.}}
 	\label{fig:resLCDM}
\end{figure*}

\begin{table*}[t!] 
 	\centering

\scalebox{1.1}{
\begin{tabular}{ccccc}       
\hline\hline                                                               

Model & Parameters & BAO + R19  & BAO + CC + SC & BAO + CC + SC + R19  \\ \hline\hline 
 & $H_0 $[km/s/Mpc] & $74.08 \pm 1.31$ & $ 69.85 \pm 1.27 $ & $71.40 \pm 0.89 $ \\
&$\Omega_m $  & $0.261 \pm 0.028$ & $0.271 \pm 0.016$ & $0.267 \pm 0.017$\\
$\Lambda$CDM  & $\Omega_\Lambda$ & $0.733 \pm 0.021$ &  $0.722 \pm 0.012$ & $0.726 \pm 0.012$\\
& $r_d$ [Mpc] & $139.0 \pm 3.1$ & $ 146.1 \pm 2.2$ & $143.5 \pm 2.0$  \\
 & $r_d/r_{fid}$ & $0.97 \pm 0.019$   & $1.01 \pm 0.021$  & $0.98 \pm 0.014 $  \\
\hline
 & $H_0 $[km/s/Mpc] & $73.76 \pm 1.52 $ & $ 70.78 \pm 0.99 $ & $ 72.01 \pm 0.93 $ \\
 &$\Omega_m $  & $ 0.181 \pm 0.051$ & $  0.253 \pm 0.011$ & $  0.252 \pm 0.009 $    \\
$\Omega_k\Lambda$CDM &$\Omega_\Lambda $  & $ 0.806 \pm 0.024 $ & $0.801 \pm 0.009 $ & $0.802 \pm 0.009 $ \\
& $r_d$  [Mpc]  & $ 143.1 \pm 3.5$ & $ 145.4 \pm 2.4 $ & $ 143.1 \pm 1.7$ \\
&$\Omega_k $  & $ -0.015 \pm 0.053 $ & $ -0.076 \pm 0.017 $ & $ -0.076 \pm 0.012 $\\
 & $r_d/r_{fid}$  & $0.962 \pm 0.019$ & $0.988 \pm 0.019 $& $0.969 \pm 0.015 $\\
 \hline
&$H_0 $[km/s/Mpc] & $73.69 \pm 1.31 $ & $  69.94 \pm 1.08$ & $ 71.65 \pm 0.88  $ \\
 &$\Omega_m $  & $ 0.243 \pm 0.039 $ & $  0.269 \pm 0.023$ & $0.266 \pm 0.022$    \\
wCDM &$\Omega_\Lambda $  & $ 0.746 \pm 0.029 $ & $  0.724 \pm 0.019$ & $0.727 \pm 0.019$ \\
& $r_d$  [Mpc] &  $138.43 \pm 3.18$ & $146.4 \pm 2.5$ & $143.2 \pm 1.9$  \\
&$w$ & $ -1.067 \pm 0.065$ & $-0.989 \pm 0.049$ & $-0.989\pm 0.049$ \\
&  $r_d/r_{fid}$  & $0.935 \pm 0.024$ & $0.99 \pm 0.0164$ & $0.967 \pm 0.015$\\

\hline\hline                                                  
\end{tabular} 
}
\caption{\it{Constraints at 95$\%$~CL errors on the cosmological parameters for the $\Lambda$CDM, $\Omega_k\Lambda$CDM model and the wCDM model. The datasets are: the BAO alone, the BAO + CC + SC combination and with with the Riess 2019 measurement as a Gaussian prior.}}
\label{tab:res}                                             
\end{table*}

\section{The standard model} 
\label{sec:LCDM}
In this model, we vary 5 parameters: $\{H_0, \Omega_m, \Omega_\Lambda, r_d, r_d/r_{d,fid}\}$. On Fig. \ref{fig:resLCDM} we report the $68\%$ and $95\%$ confidence levels for the posterior distribution of some of the parameters of the Standard $\Lambda$CDM model. The numerical values are reported in Table \ref{tab:res}. When the fit combines the {\bf CC} sample and the {\bf BAO}, the predicted values are constrained dramatically and give values closer to the one announced by Planck  (\cite{Aghanim:2018eyx}). When one adds the Riess 2019 prior for $H_0$, the fit gives a result much closer to the observed in the Supernovae sample (\cite{Riess:2019cxk}). We note that the matter energy density is smaller than the one reported by Planck ($\Omega_m^{Planck}=0.315\pm 0.007$), but this has been seen in other studies as well (\cite{Nunes:2020uex,Nunes:2020hzy}).

\begin{figure}[t!]
 	\centering
\includegraphics[width=0.45\textwidth]{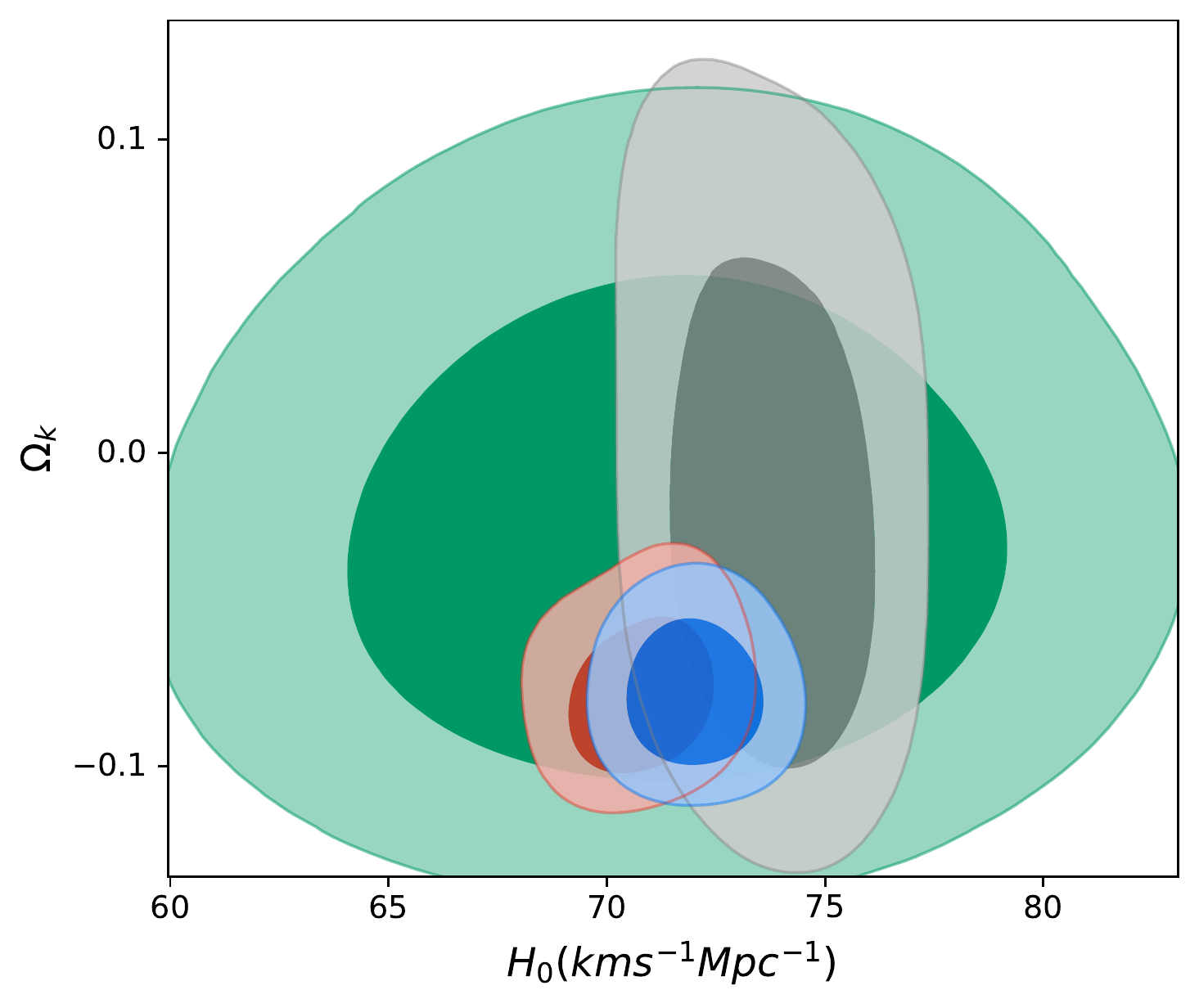}
\includegraphics[width=0.45\textwidth]{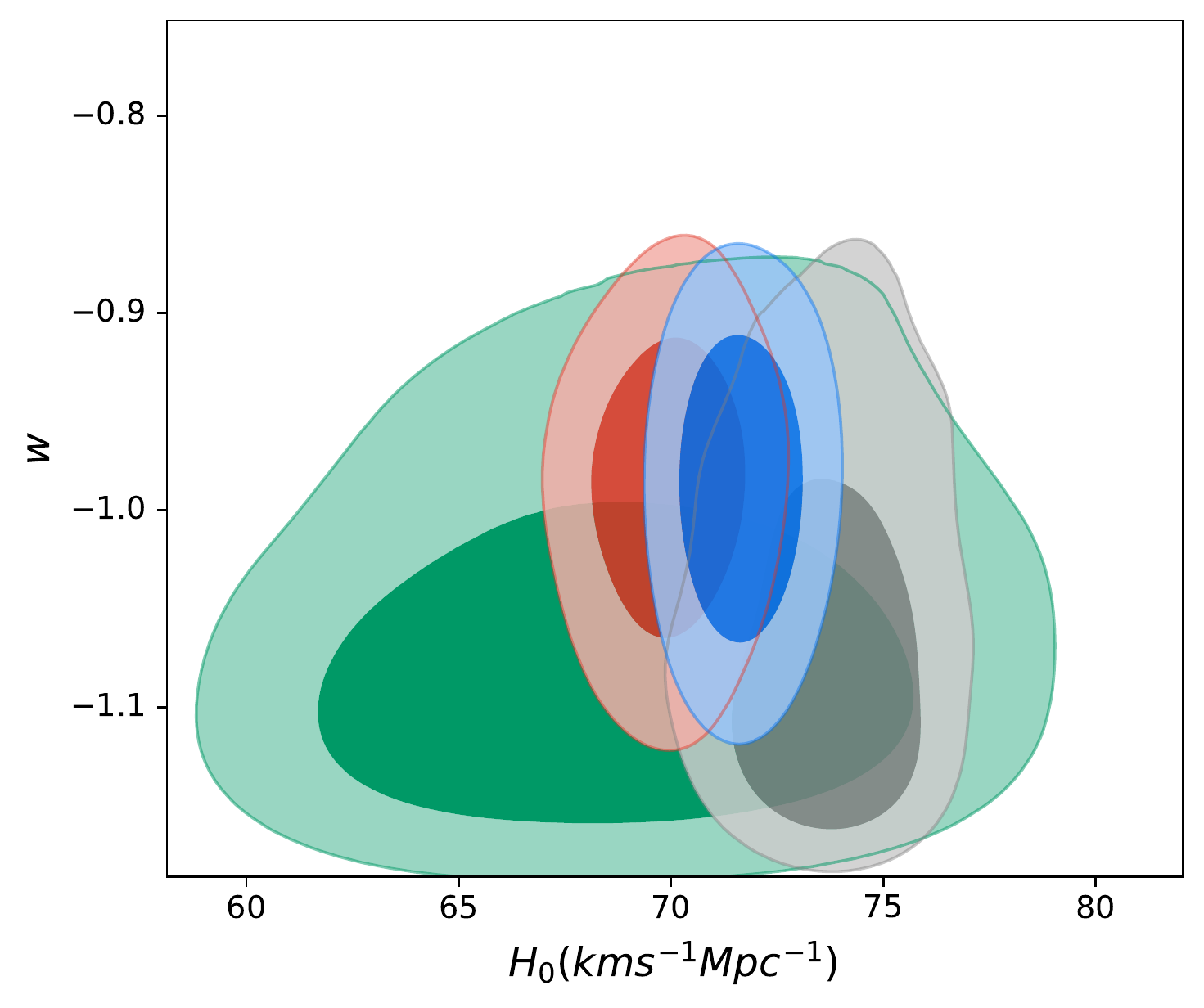}
\caption{\it{The posterior distribution for wCDM and $\Omega_k\Lambda$CDM. The upper panel shows the $\Omega_k$ vs. the $H_0$ contour and the lower panel shows the $w$ vs. the $H_0$ contour. The legend is the same as in Fig. \ref{fig:resLCDM}. }}
 	\label{fig:Okw}
 	\end{figure}

The BAO scale is set by the sound horizon at the drag epoch $z_d$ when photons and baryons decouple, given by:
\begin{equation}
r_d = \int_{z_d}^{\infty} \frac{c_s(z)}{H(z)} dz    
\end{equation}
where $c_s \approx c \left(3 + 9\rho_b /(4\rho_\gamma) \right)^{-0.5}$ is the speed of sound in the baryon-photon fluid with the baryon $\rho_b(z)$ and the photon $\rho_\gamma(z)$ densities respectively (\cite{Aubourg:2014yra}). The drag epoch corresponds to the time when the baryons decouple from the photons at  {$z_d \approx 1060$}. For a flat $\Lambda$CDM, the Planck measurements yield $147.09 \pm 0.26 Mpc$ and the WMAP fit gives $152.99 \pm 0.97 Mpc$ (\cite{Aghanim:2018eyx}). From large-scale structure combined with CC, SNea and $150.0 \pm 1.89 Mpc$ and with the local $H_0$ measurement $r_d = 143.9 \pm 3.1 Mpc$ (\cite{Aghanim:2018eyx}). Final measurements from the completed SDSS lineage of experiments in large-scale structure provide $r_d = 149.3 \pm 2.8 Mpc$ (\cite{Alam:2020sor}). Using BAO, SNea, ages of early type galaxies and local determinations of the Hubble constant,   \cite{Verde:2016ccp} reports $r_d = 143.9\pm3.1 $ Mpc. \cite{Nunes:2020uex,Nunes:2020hzy} give similar values: $144.1^{+5.3}_{-5.5} Mpc$ from $r_d/D_M$ + BBN + H0LiCOW, $150.4^{+2.7}_{-3.3} Mpc$ from $r_d/D_M$ + BBN + CC. We see that there is a big discrepancy between results from the early universe and those from the late one, similar to the $H_0$ tension. Due to this, \cite{Knox_2020} refers to it as "tensions in the $r_{d}\,-\,H_0$ plane". 

 The posterior distribution of the $r_d$ vs the Hubble parameter is presented on Fig. \ref{fig:resLCDM}. The fit for BAO + R19 yields: $139.0 \pm 3.1 Mpc$. Adding the CC and the  SC gives $146.1 \pm 2.2 Mpc $ and taking all the datasets plus the Riess prior leads to $r_d = 143.5 \pm 2.0 Mpc$. It is important to note that here the results depend critically on the prior for $r_d$ and $H_0$ as wider or narrower priors would move the preferred values in the $r_d-H_0$ plane. We see that only when working with a uniform prior our results are close to the Planck results and those of the SDSS experiments and also to earlier works taking into account only BAO data points (\cite{Zhang:2020uan}).

\section{Extensions}
\label{sec:ext}
 {We examine two types of extensions of the Standard $\Lambda$CDM model: the $\Omega_K\Lambda$CDM model and the wCDM model. For the $\Omega_k$CDM, we use as priors $\Omega_k \in [-0.1;0.1]$ and $\Omega_m \in [0.1;1-\Omega_\Lambda]$, while for the $wCDM$ we use as a prior $w \in [-1.25;-0.75]$. All the other priors are the same as for $\Lambda$CDM. The results for which can bee seen on Fig. \ref{fig:Okw} and the values can be found in Table \ref{tab:res}. }

{\it{$\Omega_k\Lambda$CDM:}} For all the 3 samples we get a negative spatial curvature energy density ($\Omega_k<0$) which corresponds to $k=1$, i.e. a closed universe. This is in line with previous results obtained by the Planck 2018 collaboration (\cite{Aghanim:2018eyx}) for CMB alone which found a preference for a closed universe at $3.4\sigma$ and also with those obtained by \cite{Li:2019qic} which includes the data from CC, Pantheon and BAO measurements to conclude also negative $\Omega_k$ for relieving the $H_0$ tension. Under the priors we use and with the full dataset, we do not observe the described tension between Planck and BAO data (\cite{DiValentino:2019qzk}). The numbers are very close to zero so they do not exclude a flat universe. From Fig.\ref{fig:Okw} is clear that the data from BAO alone has a very slight preference for a closed universe but with a big error. A smaller prior, for example $\Omega_K \in (-0.05,0.05)$ does not change the preference for a closed universe.

 {One may note the small value of $\Omega_m$ in Table \ref{tab:res} in the BAO + R19 case compared to the other cases and to the result obtained by Planck 2018:   $(\Omega_m, H_0, \Omega_k) = (0.349, 63.6, -0.011)$ (\cite{Aghanim:2018eyx}). We attribute it to the wider prior we use for $\Omega_k \in [-0.1, 0.1]$. If one is to use a tighter prior, for example $\Omega_k \in [-0.01, 0.01]$, one will get $\Omega_m=0.212\pm 0.026$, $\Omega_\Lambda=0.776\pm0.025$ with $\Omega_k=-0.009\pm 0.00008$, $H_0=73.78\pm 1.27$ and $r_d=141.5\pm2.9$. While this result is closer to the Planck results, we consider that using such a tight prior is forcing the model towards a flat universe. Similar results for a wider prior can be seen in (\cite{divalentino2020interacting}) where the matter density also fluctuates in a big interval $\Omega_m \in (0.084..0.48)$. Once again, one may explain these fluctuations with the non-trivial connection between $H_0, r_d$ and $\Omega_m$. Explicitly, increasing $\Omega_m$ leads to a smaller $r_d$, which also affects the value of $H_0$. Comparing the results for the two priors of $\Omega_k$ we see that they are consistent with Fig. 1 in \cite{Knox_2020}.}

{\it{The wCDM model:}} The dark energy equation of state we obtain differs from the one obtained by the Planck collaboration 2018 (\cite{Aghanim:2018eyx}) which gives $w=-1.03\pm 0.03$, i.e. it is  essentially consistent with a cosmological constant. In our case, it is much closer to the analysis done in (\cite{DiValentino:2020hov,Vagnozzi:2020dfn,Vagnozzi:2020zrh}) but adding the full dataset of different astrophysical objects does not exclude $w=-1$. The BAO + R19 dataset seems to tend to $w<-1$, while the full dataset seem to tend to $w\ge -1$. The BAO+R19 data is sensitive to the choice of prior so that in the case $w\in(-1.5,-0.5)$, one gets $w=-1.22$. This sensitivity does not apply to the full dataset for which the results are stable with respect to the choice of a prior. 

As a conclusion, testing the extended models with the chosen dataset and under the wide priors we use, seems to be very inconclusive and does not allow us to rule out $\Lambda$CDM as the default model. 

\section{Discussion}
\label{sec:Dis}
This work selects 17 uncorrelated BAO points from the largest collection of BAO data points (333 points), in order to decrease the intrinsic error in the posterior distribution  {due to possible correlations between measurements}. We perform a procedure which mocks adding random correlations to the covariance matrix and then we verify that indeed those additions do not change the resulting cosmological parameters significantly.

In addition to the suggested dataset we use uncorrelated points from different datasets: the Cosmic Chronometers (30 points), and the Type Ia supernova (40 selected points), points from the Hubble diagram for quasars (24 points) and Gamma Ray Bursts (162 points). Although the tension exists it has been alleviated: $2\sigma$ for the $H_0$.  Therefore, while our results suggest that cosmic acceleration can be deduced only from late time measurements, the tension with Planck values suggests a possibility for new physics (\cite{Pogosian:2020ded,Sekiguchi:2020teg,DiValentino:2019jae,DiValentino:2019ffd,DiValentino:2020kha,Ye:2020btb,Yang:2020zuk,DiValentino:2020kha,Benisty:2020xqm,Benisty:2020nql,Staicova:2016pfd,Ben-Dayan:2017rvr,Ben-Dayan:2014swa,Vasak:2019nmy,Benisty:2019jqz,Bengaly:2020neu,Alexander:2019wne,Alexander:2019ctv,Khosravi:2017hfi,Zwane:2017xbg,Narimani:2014zha,Afshordi:2008rd,Kazantzidis:2018jtb,Kazantzidis:2019dvk}).

 {The results depend critically on the priors for $r_d$ and $H_0$. From our numerical experiments, it is clear that using a tighter prior for $r_d$ will also move $H_0$ to a value compatible with that $r_d$, i.e. it will move it in the $r_d-H_0$ plane. This means that the effect from the choice of a prior for $r_d$ may be as strong as using a tight prior such as \textbf{R19} on $H_0$. On the other hand, the tight prior on $H_0$ combined with a large interval for $r_d$ essentially swings the results towards the $H_0$ measurement on which the prior is centered. We see that only when working with a uniform prior our results are close to those obtained by Planck and the SDSS experiments as in this case, our priors impose the least assumptions on the model.
}

 {The dependence on the priors is strongest in the $\Omega_k$CDM model which yields $\Omega_m=0.181\pm0.051$ for the BAO+R19 for a wider prior and $\Omega_m=0.212\pm 0.026$, for the same dataset with a tighter prior for $\Omega_k$. The Hubble parameter $H_0$ in both cases remains similar (due to the tight prior), while $r_d$ changes as explained above. The curvature energy density remains small but negative, i.e. the preference for a closed universe does not depend on the priors we have tested. .
}

{To compare both the $\Omega_k$CDM and the $w$CDM models to $\Lambda$CDM we take into account that $\Lambda$CDM is a nested model to both of the suggested extension, with 1 degree of freedom difference, on which we can use standard statistical tests. We calculate them for the total dataset without the Riess prior. For it, one can define the reduced chi-square statistic, $\chi^2_{red}=\chi^2/Dof$, where $Dof$ is the degrees of freedom of the model and $\chi^2$ is the weighted sum of squared deviations. Under the same number of runs, for the 3 models, it is close to 1 ($\chi^2/Dof_{\Lambda CDM}, \chi^2/Dof_{w}, \chi^2/Dof_{\Omega_k}\sim \{0.949,0.989,0.974\}$.  The Akaike information criteria (AIC) (\cite{AIC1,Liddle:2007fy,Anagnostopoulos:2019miu}) for a large number of data points (in our case $N=273$) can be defined as:}
\begin{equation}
 AIC=\chi_{min}^2+2p   
\end{equation}
 {where $p$ is the number of parameters of the model. Then one can calculate the AIC for the main model and its extensions. We find for $\Lambda$CDM, $w$CDM and $\Omega_k$CDM respectively: $AIC=\{264.5,266.8,272.1\}$. Since $\Lambda$CDM has the least AIC, it is the best model under this criteria. The difference between $\Lambda$CDM and the two extensions are $2.3,7.6$ AIC units respectively. This points that according to our calculations, there is some support in favor of $w$CDM and weak support in favor of  $\Omega_k$CDM \cite{Arevalo_2017}.}

Finally, comparing our results with the eBoss collaboration official results (\cite{Alam:2020sor}) we see that our results are consistent with them, even though we use a different combination of uncorrelated points for our analysis. Interestingly, we see that in our case $w\ge-1$ for the full dataset, which differs from the estimations in some of the different cases  considered in \cite{Alam:2020sor}. Our dataset differs from theirs by the inclusion of the quasars and the GRB data and the exclusion of Planck points.  {In conclusion, the BAO collection offered here, is a powerful tool for studying the cosmological parameters when combined with other estimations for the Hubble constant.}

\section*{Public Code}
The python files with the dataset and the fit package can be found in \url{https://github.com/benidav/BAO-2020}.

\begin{acknowledgements}
We thank to Adam Riess, Eduardo Guendelman, Sunny Vagnozzy, Eleonora Di Valentino and Horst Stocker for fruitful discussions. We would also like to thank the anonymous referee for their helpful comments regarding the manuscript. D.B. gratefully acknowledge the support from Frankfurt Institute for Advanced Studies (FIAS) as well the to support from Ben-Gurion University in Beer-Sheva, Israel. D.S. is thankful to Bulgarian National Science Fund for support via research grants DN 08-17, DN-18/17, KP-06-N 8/11. We have received partial support from European COST actions CA15117 and CA18108. 

\end{acknowledgements}
\bibliographystyle{aa}
%
\bibliography{ref}


\end{document}